          [2001/04/25 1.1 (PWD)]
\documentclass[a4paper,twocolumn]{esapub} 
\usepackage{graphicx}

\usepackage{times}
\usepackage{natbib}

\title{Study of Unidentified EGRET sources with INTEGRAL: first results and future prospects}

\author[1]{G. Di Cocco} 
\author[2]{A.J. Castro--Tirado}
\author[3,4]{S. Chaty}
\author[5]{A.J. Dean}
\author[6]{M. Del Santo}
\author[1]{L. Foschini}
\author[7]{N. Gehrels}
\author[1]{P. Grandi}
\author[3,4]{I. Grenier}
\author[8]{W. Hermsen}
\author[8]{L. Kuiper}
\author[9]{N. Lund}
\author[1]{G. Malaguti}
\author[3]{F. Mirabel}
\author[10]{R. Much}

\affil[1]{IASF/CNR, Sezione di Bologna, via P. Gobetti 101, 40129 Bologna, Italy}

\affil[2]{Instituto de Astrof\`{\i}sica de Andaluc\`{\i}a (IAA--CSIC), P.O. Box 03004,
E--18080 Granada, Spain}

\affil[3]{Service d'Astrophysique, Centre d'\`Etudes de Saclay, DSM/DAPNIA/SAp, CEA--Saclay,
F-91191, Gif-sur-Yvette, Cedex, France}

\affil[4]{Universit\'e Paris 7, Denis-Diderot 2, place Jussieu 75251 Paris Cedex 05, France}

\affil[5]{Department of Physics and Astronomy, Southampton University, UK}

\affil[6]{IASF/CNR, Sezione di Roma, via del Fosso del Cavaliere 100, Roma, Italy}

\affil[7]{NASA Goddard Space Flight Center, Laboratory for High Energy Astrophysics, Code
661, Greenbelt, MD 20771, USA}

\affil[8]{SRON National Institute for Space Research, Sorbonnelaan 2, 3584 CA Utrecht, The
Netherlands}

\affil[9]{Danish Space Research Institute, Juliane Maries Vej 30, 2100 Copenhagen, Denmark }

\affil[10]{ESA/ESTEC, Keplerlaan 1, Postbus 299, 2200 AG Noordwijk, The Netherlands}

\begin{document}


\maketitle

\begin{abstract}
The primary objective of the study of unidentified EGRET sources with INTEGRAL is to locate
with a few arcminute accuracy the hard X-ray / soft $\gamma$-ray counterparts within the 
EGRET error circle or to
determine upper limits on their emission in the INTEGRAL energy range. The INTEGRAL Galactic
Plane Scan (GPS) and Galactic Centre Deep Exposure (GCDE) data have been analysed and
cross--correlated with the EGRET $3^{\rm rd}$ Catalogue to search for counterparts of EGRET
sources. 
The IBIS detection of a source within the EGRET error circle of 3EG J$1736-2908$ is presented 
and its possible identification with the active galaxy GRS $1734-292$ is discussed.
Finally, preliminary results on the cross-correlation between EGRET unidentified sources and 
the first data from the IBIS survey of the Galactic Centre are presented.

\end{abstract}

\section{Introduction}

The $3^{\rm rd}$ Energetic Gamma--Ray Experiment Telescope (EGRET) Catalogue of high--energy
(E$>$100 MeV) $\gamma$--ray sources (Hartman et al. 1999)  refers to the observations carried
out by the EGRET experiment onboard the Compton Gamma--Ray Observatory (CGRO) between April
1991 and October 1995. It contains 271 point sources, including: 1 solar flare, the Large
Magellanic Cloud, 5 pulsars, the radio galaxy Centaurus A, and 66 high--confidence
identifications of blazars. In addition, 27 lower--confidence blazar identifications are
included in the catalog, which leaves 170 sources unidentified.  Since the publication of the
catalog, a lot of effort has been dedicated to the identification of the 170 unidentified
EGRET sources (Grenier 2003). While most of the extragalactic counterparts are blazars, two
possible identifications of EGRET sources with radiogalaxies have been recently proposed
(Combi et al. 2003, Mukherjee et al. 2003), in addition to the well known radiogalaxy Cen A.  
However, the multiwavelength counterpart search is hampered by the limited EGRET point spread
function, which gives error radii of $0.5^\circ-1^\circ$.

One of the two main telescopes on board INTEGRAL (Winkler et al. 2003) is the IBIS imager
(Ubertini et al. 2003).  IBIS combines for the first time broad band high energy coverage
($15$~keV -- $10$~MeV) with imaging capability.  IBIS has a large field of view
($19^\circ\times 19^\circ$ at half response), and a fine angular resolution of $12'$, sampled
in $5'$ pixels in the low energy ($0.015-1$~MeV) layer ISGRI (Lebrun et al. 2003), and in
$10'$ pixels in the high energy ($0.175-10$~MeV) detector PICsIT (Di Cocco et al. 2003).  
ISGRI reaches a point source location accuracy better than $1'$ for a $30\sigma$ detection
(Gros et al. 2003). These unprecedented characteristics for a $\gamma$--ray telescope clearly
make IBIS one of the best instruments available to date to search for counterparts of
unidentified EGRET sources. Within the framework of the Core Programme (CP) observations, a
project has been set up to systematically search for associations between INTEGRAL sources
and EGRET unidentified error sources. A first possible identification of an EGRET source has
already been found (see Di Cocco et al. 2004, A\&A submitted, for details).

\section{The case of 3EG~J$1736-2908$}

Every single pointing (Science Window, ScW) of the INTEGRAL Core Programme observations is progressively analysed, 
and all the sources detected with signal--to--noise ratio (SNR) greater than $6\sigma$ are cross--correlated 
with the $3^{\rm rd}$ EGRET Catalogue to look for possible identifications. All the INTEGRAL data analysis described 
in the present work has been done with the latest version of the Offline Standard Analysis (OSA) available 
through the INTEGRAL Science Data Centre (ISDC\footnote{\texttt{http://isdc.unige.ch/index.cgi?Soft+download}})
and whose algorithms are described in Goldwurm et al.~(2003).

\begin{figure}[!ht]
\centering
\vspace*{0.2cm}
\includegraphics[scale=0.45]{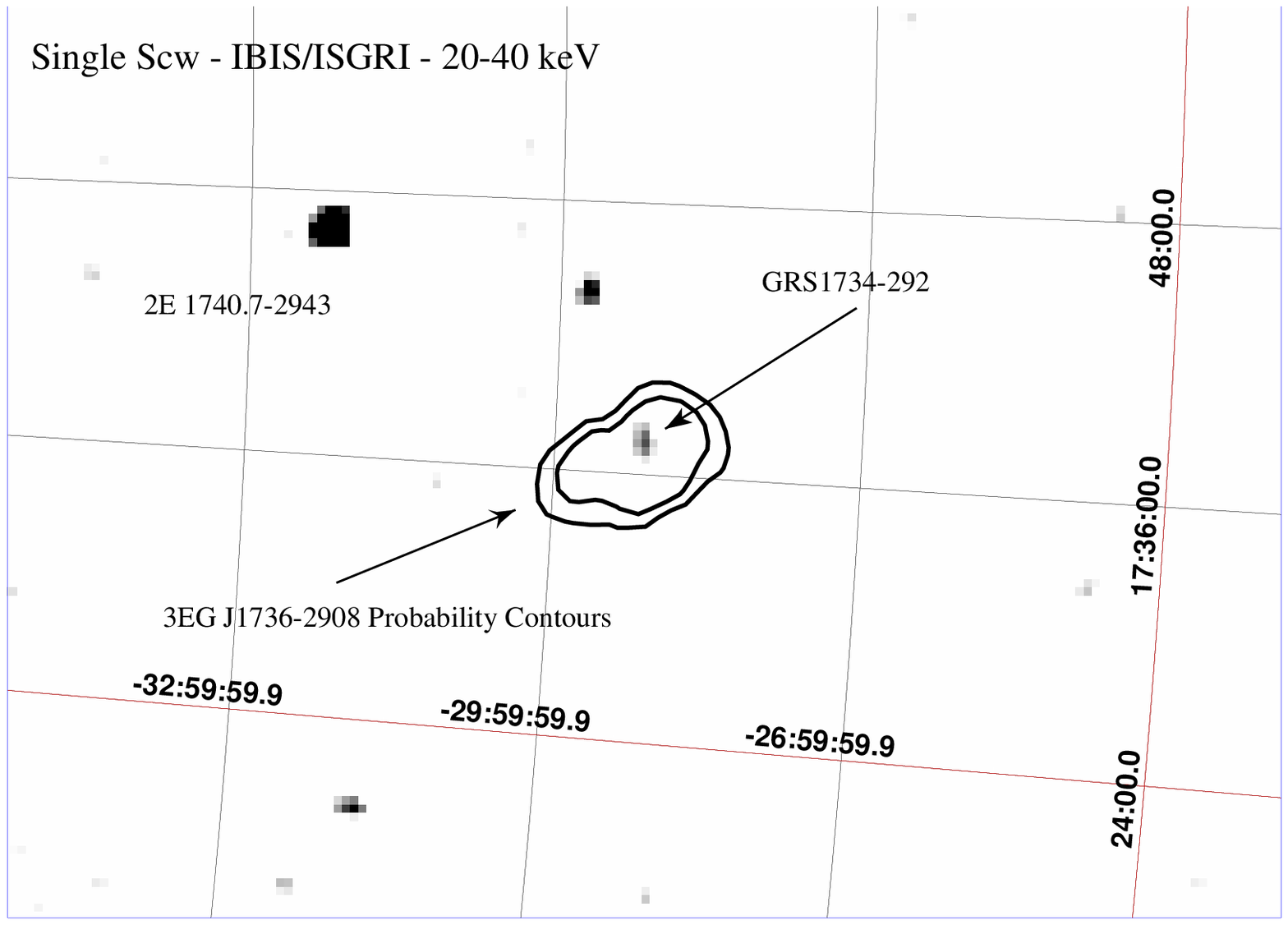}\\
\vspace*{0.2cm}
\includegraphics[scale=0.45]{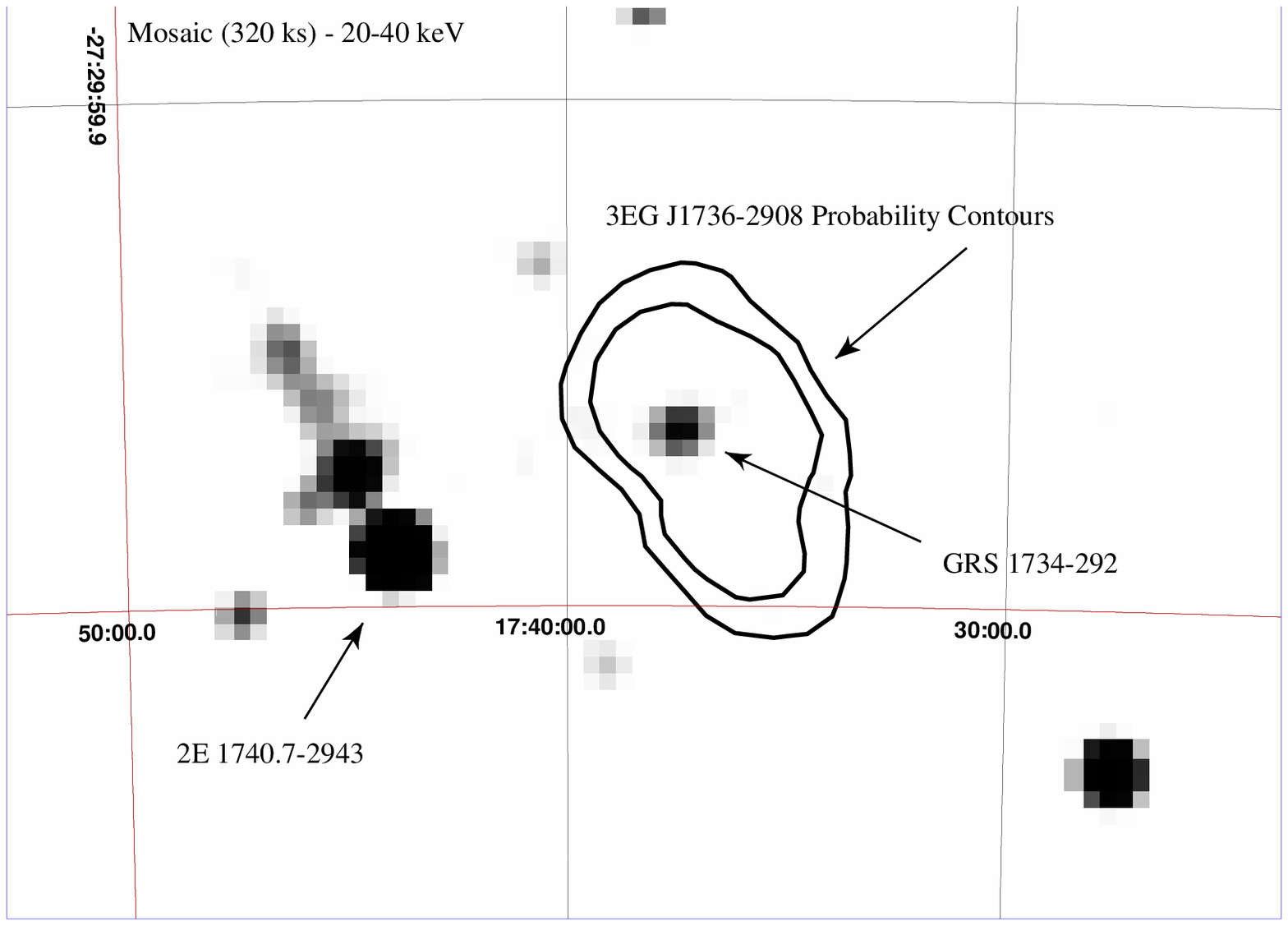}\\
\vspace*{0.2cm}
\includegraphics[scale=0.45]{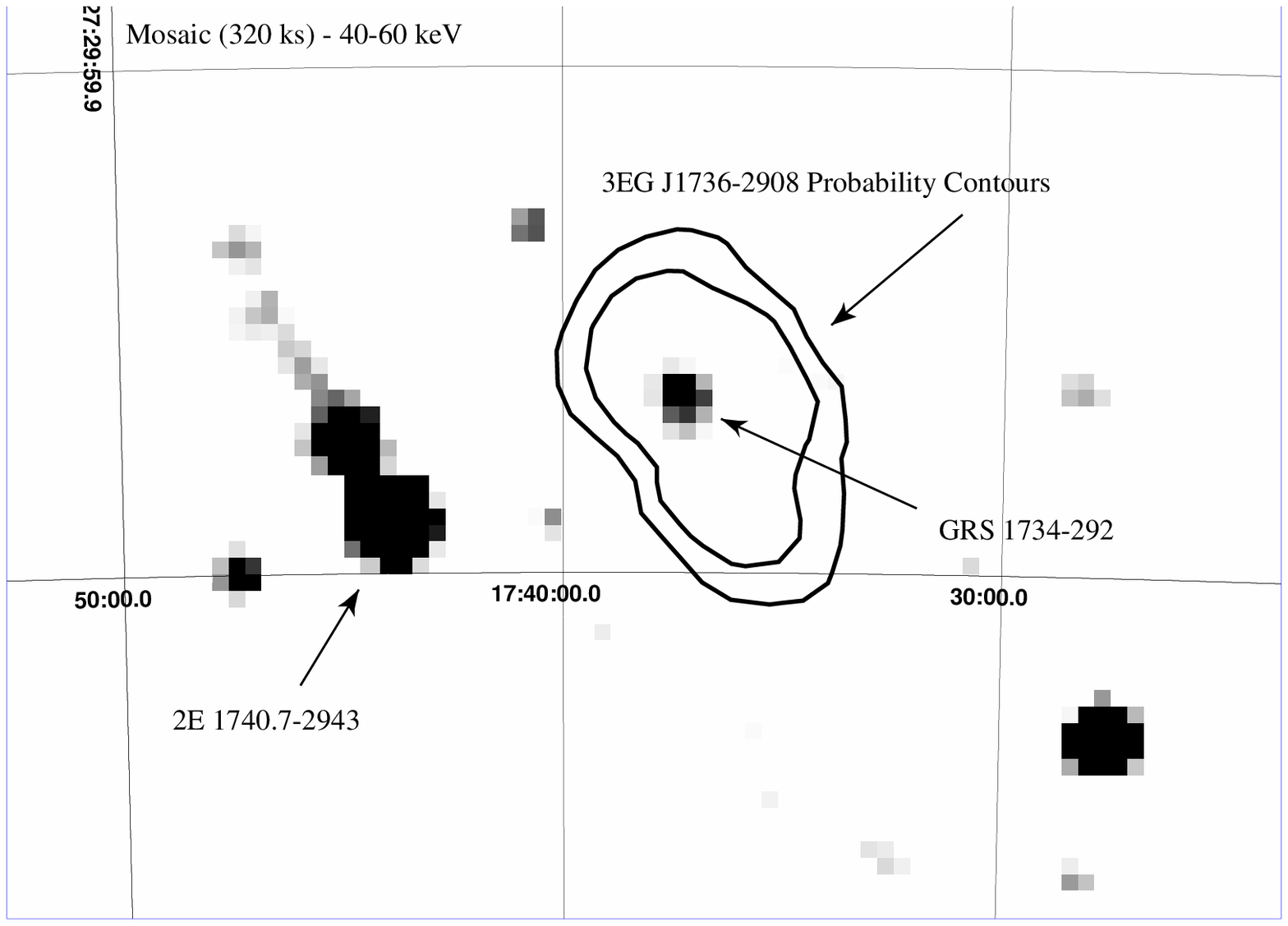}\\
\caption{(\emph{top panel:}) IBIS/ISGRI image of the 3EG J$1736-2908$ region obtained in one 
ScW (ScW 57 of Rev. 61).
(\emph{centre panel:}) IBIS/ISGRI mosaic ($320$~ks total exposure time) of the 3EG 
J$1736-2908$ region in
the energy band $20-40$~keV. (\emph{bottom panel}) The same as above, but in the energy band 
$40-60$~keV. 
The $95$\% and $99$\% EGRET probability contours of 3EG $1736-2908$ are superimposed.
(N.B.: The rotation of the figures in the centre and bottom panels with respect to the single ScW observation of the top 
panel is caused by the mosaicing procedure.)}
\label{fig:scwdet}
\end{figure}

During these scans, one source within the probability contours of 3EG J$1736-2908$ 
was found with
a statistical significance of $\approx 10\sigma$ in the ScW 57 of revolution 61 
(Fig.~\ref{fig:scwdet}, \emph{top}). The
source, later identified with the active galaxy GRS$1734-292$, had a flux of
($2.1\pm0.3$)$\times10^{-10}$ erg cm$^{-2}$s$^{-1}$ (or $28\pm4$ mCrab)  in the $20-40$~keV
energy band.

The detection significance was increased with the mosaic obtained by adding together all the ScWs
containing the 3G J$1736-2908$ region within $10^\circ$ from the centre of the field of view
(FOV). The obtained images (Fig.~\ref{fig:scwdet}, \emph{centre} and \emph{bottom} panels) 
correspond to
an effective exposure time of $\sim320$ ks. The source position ($90$\% confidence radius =
1.2$'$) obtained with this longer exposure, $\alpha=17:37:27$ and $\delta=-29:08:24$ (J2000),
is fully compatible with the single ScW detection, while the detection significances 
increased to
$17\sigma$ and $8.5\sigma$, in the $20-40$ and $40-60$ keV, respectively. The corresponding
fluxes are ($4.9\pm0.4$)$\times10^{-11}$ erg~cm$^{-2}$~s$^{-1}$ and
($2.1\pm0.3$)$\times10^{-11}$~erg~cm$^{-2}$~s$^{-1}$, respectively.

GRS$1734-292$ is also clearly visible in the mosaics of the Galactic Centre presented by Paizis et
al.~(2003), in the field of view of the black hole candidate H$1743-322$ observation by Parmar
et al.~(2003), and in the IBIS survey of the Galactic Centre region performed by Revnivtsev et
al. (2004). The flux and the detections are consistent with the present results.

For more details about the possible identification of GRS$1734-292$ as counterpart
of 3EG~J$1736-2908$ we refer the reader to Di Cocco et al. (submitted to A\&A).

\section{Search for other counterparts of EGRET unidentified sources}

\begin{figure*}[!ht]
\centering
\includegraphics[scale=0.5,angle=90]{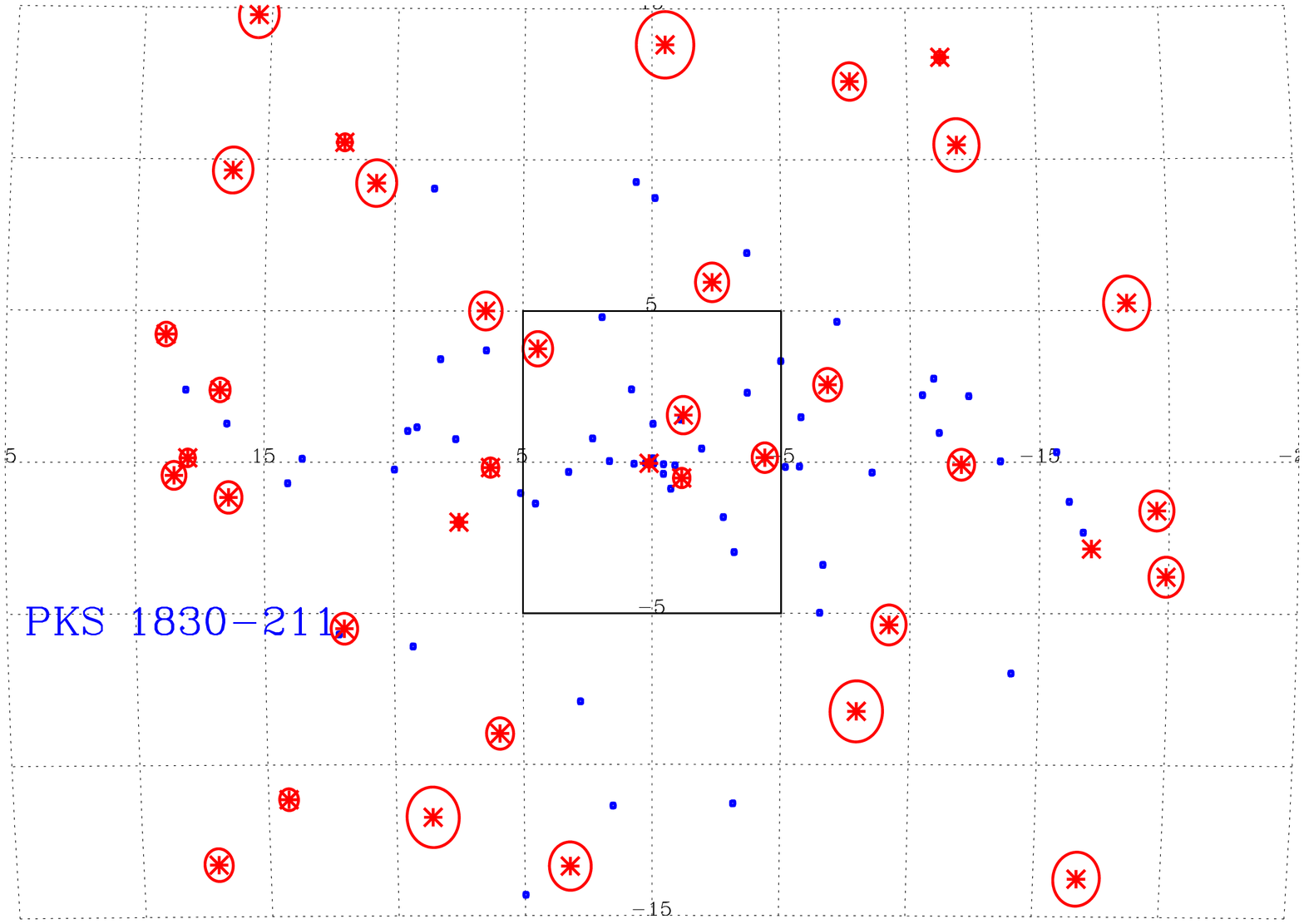}\\
\includegraphics[scale=0.5,angle=90]{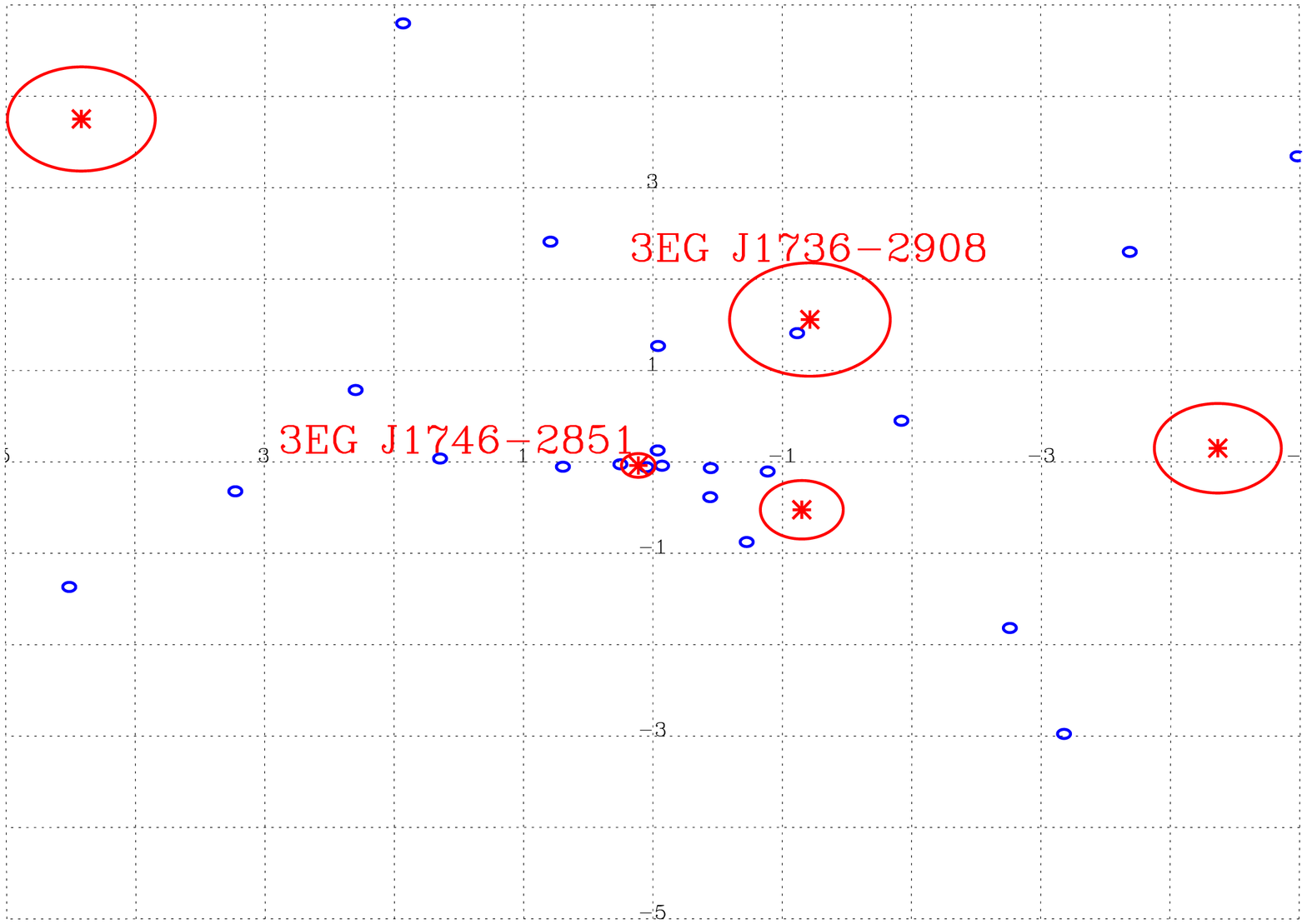}
\caption{ (\emph{Top panel:}) IBIS/ISGRI sources detected by Revnivtsev et al. (2004) in the 
Galactic Centre region
survey ($3'$ radius, blue circles) superimposed to the EGRET error circles (95\% probability radius, red asterisks) in the 
same
region. (\emph{Bottom panel:}) The zoom panel clearly shows the coincidence 
position of the source
identified as GRS$1734-292$ with 3EG J$1736-2908$.}
\label{fig:mike}
\end{figure*}

The 2~Ms IBIS/ISGRI mosaic by Revnivtsev et al.~(2004) gives a list of sources detected in the central radian of the Milky Way
down to a limiting flux of $1.3\times10^{-11}$~erg~cm$^{-2}$~s$^{-1}$. Therefore, we performed a cross--correlation of the
$3^{\rm rd}$ EGRET Catalog with the source list of Revnivtsev et al. (2004). The results shown in Fig.~\ref{fig:mike} confirm
that GRS$1734-292$ is the only source detected by IBIS inside the error circle of 3EG~J$1736-2908$. On the other hand, while
two IBIS sources are found within the 95\% error circle of the EGRET unidentified source 3EG J$1746-2851$ at the Galactic
Centre, no other unidentified EGRET sources match with the INTEGRAL detections. The only other EGRET source coincident with an
IBIS source is a known blazar, PKS $1830-211$ which lies within the error circle of 3EG J$1832-2110$.

\section{Conclusions}

We presented the first results of the early activity in the search for
counterparts of the unidentified EGRET sources in the framework of the
INTEGRAL Core Programme. One candidate counterpart has been found close to the
Galactic Centre and the identification with the active galaxy GRS$1734-292$ has been 
proposed.
The cross--correlation of the 2~Ms mosaic of the central radiant (Revnivstev et al. 2004)
with the $3^{\rm rd}$ EGRET catalog has not produced other possible counterparts.
The research should now be extended to the others regions of the sky covered by
the INTEGRAL core programme.

\section*{Acknowledgments}
Based on observations obtained with \emph{INTEGRAL}, an ESA project with instruments and science
data centre funded by ESA member states (especially the PI countries:
Denmark, France, Germany, Italy, Switzerland, Spain), Czech Republic and
Poland, and with the participation of Russia and the USA.

The italian participation to the INTEGRAL/IBIS project is financed by the Italian Space Agency (ASI).

\end{document}